\documentclass[showpacs,superscriptaddress,showkeys,aps,prl,twocolumn,citeautoscript]{revtex4}
\usepackage{graphicx}
\usepackage{pstricks}

\newcommand{\degr}{\(^\circ\)}
\newcommand{\ignore}[1]{}
\newcommand{\etal}{{\it et al.}}

\begin{document}

\title{Controlling polarization at insulating surfaces:\\
quasiparticle calculations for molecules adsorbed on insulator films}
\author{Christoph Freysoldt}
\affiliation{Fritz-Haber-Institut der Max-Planck-Gesellschaft,
             Faradayweg 4--6, 14195 Berlin, Germany}
\author{Patrick Rinke}
\affiliation{Fritz-Haber-Institut der Max-Planck-Gesellschaft,
             Faradayweg 4--6, 14195 Berlin, Germany}
\affiliation{Materials Department, University of California, Santa Barbara, CA 93106, USA}
\author{Matthias Scheffler}
\affiliation{Fritz-Haber-Institut der Max-Planck-Gesellschaft,
             Faradayweg 4--6, 14195 Berlin, Germany}
             
\begin{abstract}
By means of quasiparticle-energy calculations in the $G_0W_0$ approach, we show for the prototypical insulator/semiconductor system NaCl/Ge(001) that polarization effects at the interfaces noticeably  affect the excitation spectrum of molecules adsorbed on the surface of the NaCl films. The magnitude of the effect can be controlled by varying the thickness of
the film, offering new opportunities for tuning electronic excitations in e.g. molecular electronics or quantum transport.
Polarization effects are visible even for the excitation spectrum of the NaCl
films themselves, which has important implications for the interpretation of surface science experiments for the characterization of  insulator surfaces.
%
\end{abstract}

\pacs{
   73.20.-r; 
   71.20.Ps; 
}
\keywords{NaCl/Ge(001), GW, polarization, thin film}

\maketitle

On the nanoscale, materials often reveal extraordinary features. 
To harness this potential it is essential to grow or manufacture nanostructures in a controlled way.
Ultrathin insulating films are an example for which this has been achieved
on metal and semiconductor surfaces.
We have recently shown that these films develop new and unique properties as their thinness approaches the limit of a few atomic layers 
and that such supported ultrathin films should be regarded as new nanosystems in their own rights \cite{oxidePRL}. 
Here we go one step further and demonstrate by means of first principles calculations that control over the film thickness means control over the polarization of the film. This in turn gives access to properties on the film's surface, for example the energy levels of molecular adsorbates, which are relevant in the context of e.g. catalysis, molecular electronics, or quantum transport.

The fact that ultrathin insulator films offer a new perspective of control at the nanoscale is increasingly being recognized. Repp \etal{}, for example,  have recently demonstrated that gold atoms adsorbed on NaCl/Cu can be reversibly switched between the neutral and negative charge state \cite{ReppScience04}. 
Alternatively, the charge state of Au atoms on MgO films can be controlled by the film thickness \cite{Sterrer}. 
For planar molecules adsorbed on Cu-supported NaCl films, the 
molecular orbitals can be resolved spatially and energetically \cite{Repp05,ReppAPL} by scanning tunneling microscopy (STM) and spectroscopy (STS),
and even reactions can be followed \cite{ReppScience06}.
Interestingly, STS experiments performed on  pentacene molecules adsorbed on NaCl/Cu(111) show
a significant influence of the film thickness on the molecular gap \cite{ReppAPL}.

Ultrathin insulator films have also developed into highly valuable and intensively studied model systems for characterizing
insulating surfaces. The study of insulator surfaces has proven difficult due to their lack of conductivity, which severely limits   
the range of applicable surface science techniques.
Ultrathin insulator films grown on conducting substrates offer a solution to this dilemma because the films can exchange
electrons with the substrate by tunneling \cite{FreundReview,GoodmanReview}.
Caution has to be applied, however, when transferring thin-film results to the surfaces of technological interest.
The properties of ultrathin films may deviate considerably from those of macroscopic films  \cite{oxidePRL} and
the excitation spectrum may be affected by  the polarization effects presented here. 

\begin{figure}
\center
\includegraphics[width=0.45\textwidth,trim=70 40 70 0,clip]{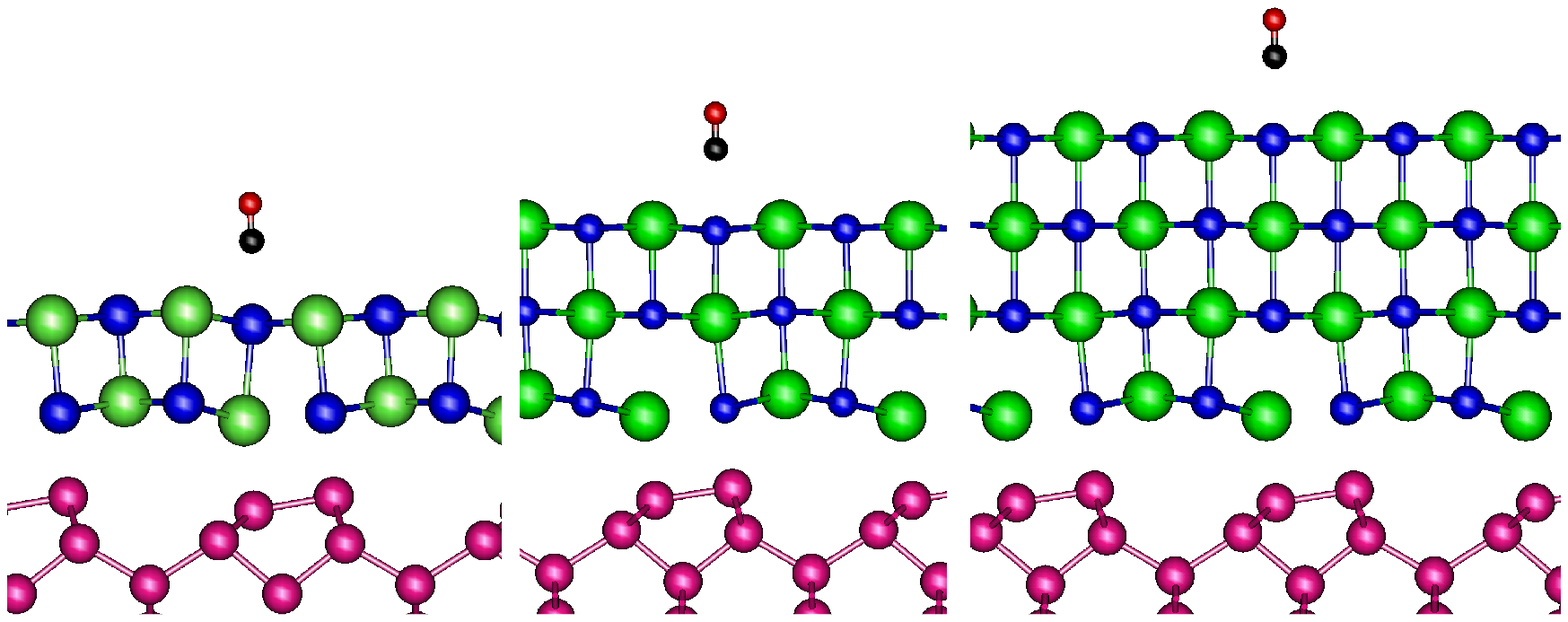}
\caption{(Color online). 
Structure of NaCl films (2--4 layers) on Ge(001) with a CO adsorbate.
The spheres denote the atoms: Ge (medium magenta/gray),
Na (large blue/dark), Cl (large green/gray), C (small black), 
O (small red/gray).
}
\label{fig:naclfilmstructure}
\end{figure}

In this Letter we will address both of these points by means of 
$G_0W_0$ quasiparticle-energy calculations \cite{Aulbur,Rinke}
for the example of a prototypical 
insulator-semiconductor interface (NaCl/Ge(001)) and CO as a model adsorbate.
Supported NaCl films are well-behaved model systems for studying the properties
of insulator surfaces \cite{Foelsch,Zielasek00,Olsson,ReppScience06}.
Although they are mostly grown on metals, notably Cu \cite{Olsson,ReppScience04}, Ge(001) is the substrate of choice
for studying insulator/semiconductor interfaces  \cite{Foelsch,Schwennicke,Gloeckler,Zielasek,Zielasek00}.
In recent years, these films have also attracted increasing interested in the context of STM and STS studies of atomic and molecular adsorbates 
 \cite{Repp05,ReppScience04,
Sterrer,ReppAPL,ReppScience06}.

We use density-functional theory (DFT) in the local-density approximation (LDA)  to determine the atomic structure of NaCl films on Ge(001). The electronic excitation spectrum is calculated with many-body perturbation theory in the $GW$ approach as perturbation to the LDA ground state (henceforth denoted $G_0W_0$@LDA). The GW approach has not only become the method of choice for calculating quasiparticle excitations in solids \cite{Aulbur,Rinke} as probed in STS or direct and inverse photoemission, but also includes long-range polarization effects. In a bulk material these encompass the screening of additional charge. At a surface or an interface, however, the abrupt change in dielectric constant gives rise to a net build up of charge, so called image charges.
This net polarization acts back on the additional charge and increases in strength with decreasing distance between the additional charge and the interface. 
These polarization effects are absent from the most common density functionals (such as the local-density or generalized gradient approximation, exact-exchange and hybrid functionals), but enter the $GW$ self-energy ($\Sigma$=$iGW$) through the screened Coulomb potential $W$. The application of the $GW$  method is therefore necessary to capture these effects \cite{White/Godby/Rieger/Needs:1997,Fratesi1,Rohlfing/Wang/Krueger/Pollmann:2003,ClusterImStates:2004}.

Polarization or image effects are present at any surface or interface, but are most commonly associated with metal surfaces, where the ratio in dielectric constants is largest. In supported ultrathin films, however, a charged excitation
(e.g. an electron added to the 2$\pi^*$ level of CO on NaCl/Ge(001)) polarizes two interfaces (vacuum/NaCl and NaCl/Ge(001)). 
The combination of dielectric constants and film thickness therefore controls the strength of the polarization effects from the semiconductor/insulator interface to the insulator surface.

Before we address the polarization effects at the NaCl/Ge interface
in detail we briefly describe its atomic structure, 
which had not been determined previously.
The DFT-LDA calculations were performed with the SFHIngX code\cite{SFHIngX}.
We employ a plane-wave basis set (40 Ry cutoff)
and norm-conserving pseudopotentials.
The NaCl/Ge system is modeled in the repeated slab-approach with a 6-layer
Ge slab at the experimental lattice constant (saturated by hydrogen atoms
on the bottom side) plus a varying 
number NaCl layers on the top side \cite{SizeComment}.
Increasing the Ge thickness to 12 layers produces no significant change
in the atomic or electronic structure of the NaCl films.
In agreement with experimental indications \cite{Zielasek},
we find that the Ge dimers of the clean Ge(001) surface prevail below
the NaCl film, giving rise to a $2\times1$ surface lattice \cite{Foelsch}.
The Ge dimers below the NaCl film remain asymmetric, but with a smaller tilting
angle (10\degr) compared to the free surface (19\degr).
The adhesion of the film is dominated
by the electrostatic interaction between the ions in the film and
the partial charges developing at the buckled-dimer surface of Ge(001).
The relaxation pattern in the bottom layer follows
the electrostatic profile of the Ge surface, which attracts
the ions next to the dimer, but repels those above the 
inter-dimer troughs (cf. Fig.~\ref{fig:naclfilmstructure}).
The corrugation in the higher layers is induced by the 
bottom layer and quickly flattens out as the thickness increases.

Next, we use the adsorption of a small molecule to probe the effect of the interface polarization at the film's surface.
For this purpose, we placed a single CO molecule in the $2\times1$
surface unit cell (we have no indications for relevant changes at
lower coverages).
CO physisorbs perpendicular to the NaCl(001) surface
with the C-end down (cf. Fig.~\ref{fig:naclfilmstructure}).
The CO axis tilts along (110), i.e. the Ge dimer, by 2.6\degr{} for 2 monolayer(ML) NaCl,
0.7\degr{} (3 ML), and 0.1\degr{} (4 ML), respectively.
The adsorption energy of 0.28\,eV is the same for the two inequivalent
Na sites at the surface and does not depend on the film's thickness
to within 0.01\,eV. It also agrees with the value for a thick free-standing
NaCl slab.
The molecular states give rise to flat bands in the band structure
and do not hybridize with NaCl states.
In the following, we focus on the molecular gap
(given by the $5\sigma$-$2\pi^*$ splitting) at the
$\overline\Gamma$ point.

The $G_0W_0$ calculations for the electron addition and removal spectra were performed with the $\texttt{gwst}$ code \cite{RiegerEtal,gwstGLEtal,gwstAnisoEtal}.
For the correlation (exchange) self-energy, a 14 Ry (28 Ry) plane-wave cutoff
and a $6\times 3\times 1$ $\mathbf k$-point sampling was used.
State summations included 2500 bands (81 eV above the Fermi level). 
To correct for artificial polarization effects in the repeated-slab approach,
a ''finite vacuum'' correction was applied \cite{slabgw}.

\begin{figure}
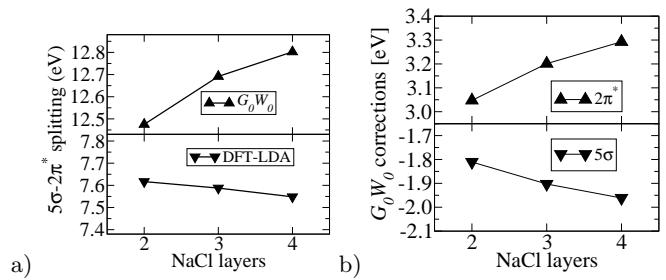

\center
a)
\includegraphics[width=0.215\textwidth,clip]{genacl-co-gap.eps}
b)
\includegraphics[width=0.215\textwidth,clip]{genacl-co-qpcorr.eps}
\caption{ a) $5\sigma$-$2\pi^*$ splitting of CO on NaCl/Ge(001) at the 
LDA and  $G_0W_0$@LDA levels of theory, and 
b) $G_0W_0$ corrections to the $5\sigma$ 
and $2\pi^*$ molecular orbitals
as a function of the NaCl film thickness.
} 
\label{fig:ge3nacl-co-gap}
\end{figure}
In Fig.~\ref{fig:ge3nacl-co-gap}, we compare the molecular gap of CO 
computed for LDA  and $G_0W_0$@LDA for films of 2--4 ML thickness. At
the level of LDA, we observe a small reduction of the molecular gap with
increasing film thickness due to the thickness-dependent structural changes
in the surface of the NaCl film.
The $G_0W_0$ corrections, however, reverse this trend and introduce
a significant increase in the quasiparticle gap from 12.48\,eV (for 2 ML)
to 12.80\,eV (4 ML).
The other CO orbitals exhibit analagous thickness-dependent shifts (not shown).
%
These gaps are significantly smaller than for CO on a pure NaCl
surface (13.1\,eV for 6 ML NaCl, no Ge) or the free molecule (15.1\,eV).
Similar reductions have been found in $G_0W_0$ calculations
for molecules adsorbed on
insulators \cite{RohlfingMgO} or semimetals \cite{benzeneGraphite} and are a result of surface polarization or charge transfer to the molecule \cite{Thygesen/Rubio:2009}.

We will now demonstrate that the reduction of the 
5$\sigma$-$2\pi^*$ splitting is caused by polarization effects
at the two interfaces.
Inspection of Fig.~\ref{fig:ge3nacl-co-gap}b) reveals that the thickness-dependent changes are of similar magnitude, but opposite sign.
Such a behavior is characteristic for long-range polarization effects
\cite{slabgw,benzeneGraphite}.
To illustrate this, we split the electron addition/removal into two steps.
First, the charged state is created
on a free molecule. Secondly, we consider the electronic polarization of
the NaCl/Ge substrate.
If the lifetime of the charged state on the molecule is long compared
to the electronic relaxation time of the substrate, we may
treat the rearrangement of charge with classical electrostatics.
The polarization lowers the energy of the charged state
$E_{N\pm1}$ by $\frac 12 q\Delta V$, where $\Delta V$ is the 
polarization-induced change in the electrostatic potential
and $q$ the charge.
The energy of the hole state ($E_N - E_{N-1}$) thereby increases
whereas that of the electron state ($E_{N+1}-E_N$) becomes smaller.

To substantiate this picture, we estimate $\Delta V$
outside the supported NaCl films.
For this, the Ge substrate, the NaCl film, and the vacuum region are
replaced by homogeneous dielectric media ($\varepsilon$=14, 2.8, and 1,
respectively) with abrupt interfaces. The thickness of the NaCl region
is taken to be 2.8\,\AA{}/layer.
The effective polarization is then computed using the image-charge method.
For $\Delta V$,  we then take the value of the image potential at the
position of the CO molecule outside the surface.
Since this position is somewhat ambiguous due to the spatial extent
of the CO orbitals we determine it by requiring that the reduction
of the $5\sigma$-$2\pi^*$ splitting of 2 eV at the bare NaCl surface
is reproduced by the model.
This yields a value of 1.5 \AA, and the model then gives
$5\sigma$-$2\pi^*$ splittings of 12.6, 12.7, and 12.8\,eV
for the 2, 3, and 4 ML films, respectively, in   
good agreement with the values from
our $G_0W_0$@LDA calculations. 

\begin{figure}
\center
\includegraphics[width=0.45\textwidth,clip]{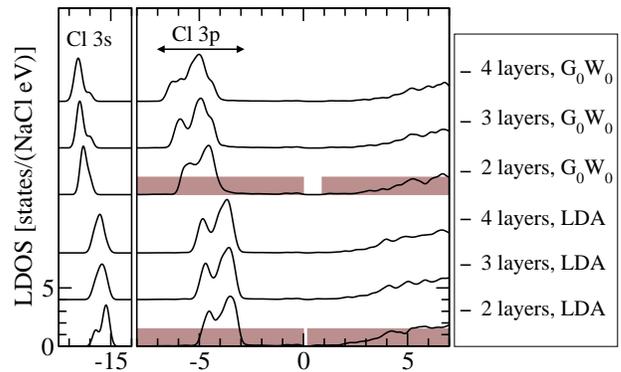}
\caption{DOS inside the supported NaCl films of 2--4 ML thickness
at the LDA and the $G_0W_0$@LDA level. The shaded region indicates
the energy position of the Ge bands.
The  valence band maximum of the Ge slab defines the energy zero.}
\label{fig:genacl-ldos}
\end{figure}

On the experimental side, a gap reduction as a function of film thickness
has been reported for STS experiments on pentacene molecules adsorbed on
NaCl/Cu(111) \cite{ReppAPL}.
In STS, the molecular states give rise to tunneling resonances and the observed tunneling gap amounts to 3.3, 4.1, and 4.4\,eV
for NaCl films of 1,2, and 3 ML in thickness (the gap of pentacene in the gas phase is 5.3 eV).
The overall trend as well as the magnitude agree very well with our
$G_0W_0$@LDA calculations for CO/NaCl/Ge considering that the ratio of dielectric constants is much larger in the NaCl/Cu case.
A tunneling resonance gap will be hard to observe experimentally for CO/NaCl/Ge, however, because the highest occupied molecular orbital (5$\sigma$) is located $\sim$8.5\,eV below the Ge valence band maximum (vbm) in $G_0W_0$@LDA and thus even below the NaCl valence band (4-7 eV below the Ge vbm, cf. 
Fig.~\ref{fig:genacl-ldos}).

Our results highlight that surface and interface polarization effects are important for adsorbed atoms and molecules. Supported ultrathin films thereby offer unprecedented opportunities for controlling these effects by tailoring both the film's thickness and dielectric constant to the desired properties. These additional parameters may expedite the design of devices in molecular electronics or quantum transport, where the distance of a molecular state to the Fermi level is an important quantity.

We will now demonstrate that the same polarization effects also affect excitations inside the ultrathin film and discuss the implications of this for the characterization of insulator surfaces.
Fig.~\ref{fig:genacl-ldos} shows the density of states (DOS) projected onto the
NaCl film for films with 2--4 ML for LDA with and without the $G_0W_0$ 
corrections.
The most important change when comparing the LDA and $G_0W_0$ DOS
is the shift of the NaCl bands \cite{footnote:VBstates} relative to the Ge states.
This is not too surprising, as the $G_0W_0$ 
corrections to the bulk band gaps are much larger for NaCl (3.3\,eV) than
for Ge (0.7\,eV).
Including the $G_0W_0$ corrections, the top of the film's valence
band lies $\sim$4.2\,eV below that of the Ge substrate in excellent
agreement with ultraviolet photoelectron spectroscopy \cite{Barjenbruch}.
More remarkable, however, is the change in the shape of the NaCl-derived features when going from LDA
to $G_0W_0$@LDA and from 2 to 4 layers. This indicates that the $G_0W_0$ shifts for the NaCl states are not uniform and that corrections derived from bulk calculations are not easily transferable to thin films. Excited states are instead subject to additional
thickness-dependent and substrate-specific changes that are neither encompassed by a ground-state perspective nor easily derivable from bulk $G_0W_0$ calculations for the separate fragments alone. These thickness-dependent variations
should be observable in high-resolution spectroscopic experiments.

In LDA the DOS of bulk NaCl (not shown) is already attained at a thickness of only three layers. A similar behavior has previously been observed for ultrathin silica, hafnia and alumina films \cite{oxidePRL}. However, this is no longer the case when charged excitations (e.g. photoemission or tunneling) are treated appropriately. The  $G_0W_0$@LDA calculations demonstrate clearly that the DOS of ultrathin films differs from that of bulk NaCl (which is identical in shape to the LDA DOS for a 4 layer film).
This implies that caution has to be applied when interpreting spectroscopic results.
Spectra of ultrathin films are not representative of bulk samples.

\begin{figure}
\center
\includegraphics[width=0.45\textwidth]{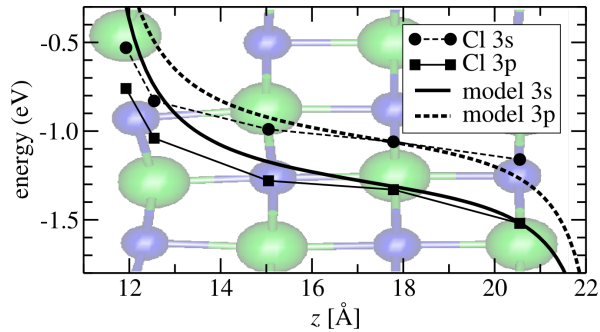}
\caption{(Color online) Comparison of the extrapolated $G_0W_0$ orbital
   shifts for the 4-layer NaCl film to the image
potential of the dielectric slab model (cf. text).}
\label{fig:qpshift-model}
\end{figure}

The non-uniform $G_0W_0$ shifts are also a result of the interface polarization effects.
The position-dependence of the self-energy ($\Sigma=iGW$) can be understood with the same dielectric slab model introduced above. Solving the image-charge model yields the induced potential $\Delta V(z)$ as a function of the position in the film $z$.
From this, we build the following model self-energy for the occupied NaCl states
\begin{equation}
\Sigma(z) = \Delta\Sigma - \frac12 \Delta V(z)
\label{eq:modelSigma}
\;.
\end{equation}
The constant $\Delta\Sigma$ encompasses all $GW$ self-energy effects that
are not due to image effects. Its  actual value is not important for
analyzing the position dependence of the self-energy.
Based on our $G_0W_0$ calculations,
we estimate it to be $-1.05\,$eV for Cl $3s$ and $-1.3\,$eV for Cl $3p$.
In order to assess the local variation of the self-energy in the full $G_0W_0$ 
calculation,
the valence bands are projected onto atomic orbitals. For states that are predominantly composed of orbitals from a single atomic layer  ($>$80\%) we observe a linear dependence between the $G_0W_0$ correction and the projection weight.  By extrapolating to 100\%, the orbital-dependent quasiparticle corrections shown in Fig.~\ref{fig:qpshift-model} are obtained. 
The variation of the quasiparticle shifts throughout the NaCl film ($G_0W_0$ DOS in Fig.~\ref{fig:genacl-ldos}) is reproduced well by a projection-weighted sum of these local-orbital contributions.
In Fig.~\ref{fig:qpshift-model}, we compare the extrapolated shifts
to our model self-energy.
The good agreement illustrates that the interface polarization effects in the Ge/NaCl system are indeed the cause for the position dependence in the $G_0W_0$ self-energy.

We acknowledge fruitful discussions with Philipp Eggert 
and the Nanoquanta Network of Excellence (NMP4-CT-2004-500198) for financial
support.


\begin{thebibliography}{31}
\expandafter\ifx\csname natexlab\endcsname\relax\def\natexlab#1{#1}\fi
\expandafter\ifx\csname bibnamefont\endcsname\relax
  \def\bibnamefont#1{#1}\fi
\expandafter\ifx\csname bibfnamefont\endcsname\relax
  \def\bibfnamefont#1{#1}\fi
\expandafter\ifx\csname citenamefont\endcsname\relax
  \def\citenamefont#1{#1}\fi
\expandafter\ifx\csname url\endcsname\relax
  \def\url#1{\texttt{#1}}\fi
\expandafter\ifx\csname urlprefix\endcsname\relax\def\urlprefix{URL }\fi
\providecommand{\bibinfo}[2]{#2}
\providecommand{\eprint}[2][]{\url{#2}}

\bibitem[{\citenamefont{Freysoldt
  et~al.}(2007{\natexlab{a}})\citenamefont{Freysoldt, Rinke, and
  Scheffler}}]{oxidePRL}
\bibinfo{author}{\bibfnamefont{C.}~\bibnamefont{Freysoldt}},
  \bibinfo{author}{\bibfnamefont{P.}~\bibnamefont{Rinke}}, \bibnamefont{and}
  \bibinfo{author}{\bibfnamefont{M.}~\bibnamefont{Scheffler}},
  \bibinfo{journal}{Phys. Rev. Lett.} \textbf{\bibinfo{volume}{99}},
  \bibinfo{pages}{086101} (\bibinfo{year}{2007}{\natexlab{a}}).

\bibitem[{\citenamefont{Repp et~al.}(2004)\citenamefont{Repp, Meyer, Olsson,
  and Persson}}]{ReppScience04}
\bibinfo{author}{\bibfnamefont{J.}~\bibnamefont{Repp}},
  \bibinfo{author}{\bibfnamefont{G.}~\bibnamefont{Meyer}},
  \bibinfo{author}{\bibfnamefont{F.~E.} \bibnamefont{Olsson}},
  \bibnamefont{and} \bibinfo{author}{\bibfnamefont{M.}~\bibnamefont{Persson}},
  \bibinfo{journal}{Science} \textbf{\bibinfo{volume}{305}},
  \bibinfo{pages}{493} (\bibinfo{year}{2004}).

\bibitem[{\citenamefont{Sterrer et~al.}(2007)\citenamefont{Sterrer, Risse,
  Pozzoni, Giordano, Heyde, Rust, Pacchioni, and Freund}}]{Sterrer}
\bibinfo{author}{\bibfnamefont{M.}~\bibnamefont{Sterrer}},
  \bibinfo{author}{\bibfnamefont{T.}~\bibnamefont{Risse}},
  \bibinfo{author}{\bibfnamefont{U.~M.} \bibnamefont{Pozzoni}},
  \bibinfo{author}{\bibfnamefont{L.}~\bibnamefont{Giordano}},
  \bibinfo{author}{\bibfnamefont{M.}~\bibnamefont{Heyde}},
  \bibinfo{author}{\bibfnamefont{H.-P.} \bibnamefont{Rust}},
  \bibinfo{author}{\bibfnamefont{G.}~\bibnamefont{Pacchioni}},
  \bibnamefont{and} \bibinfo{author}{\bibfnamefont{H.-J.}
  \bibnamefont{Freund}}, \bibinfo{journal}{Phys. Rev. Lett.}
  \textbf{\bibinfo{volume}{98}}, \bibinfo{pages}{096107}
  (\bibinfo{year}{2007}).

\bibitem[{\citenamefont{Repp et~al.}(2005)\citenamefont{Repp, Meyer,
  Stojkovi\'c, Gourdon, and Joachim}}]{Repp05}
\bibinfo{author}{\bibfnamefont{J.}~\bibnamefont{Repp}},
  \bibinfo{author}{\bibfnamefont{G.}~\bibnamefont{Meyer}},
  \bibinfo{author}{\bibfnamefont{S.~M.} \bibnamefont{Stojkovi\'c}},
  \bibinfo{author}{\bibfnamefont{A.}~\bibnamefont{Gourdon}}, \bibnamefont{and}
  \bibinfo{author}{\bibfnamefont{C.}~\bibnamefont{Joachim}},
  \bibinfo{journal}{Phys. Rev. Lett.} \textbf{\bibinfo{volume}{94}},
  \bibinfo{pages}{026803} (\bibinfo{year}{2005}).

\bibitem[{\citenamefont{Repp and Meyer}(2006)}]{ReppAPL}
\bibinfo{author}{\bibfnamefont{J.}~\bibnamefont{Repp}} \bibnamefont{and}
  \bibinfo{author}{\bibfnamefont{G.}~\bibnamefont{Meyer}},
  \bibinfo{journal}{Appl. Phys. A} \textbf{\bibinfo{volume}{85}},
  \bibinfo{pages}{399} (\bibinfo{year}{2006}).

\bibitem[{\citenamefont{Repp et~al.}(2006)\citenamefont{Repp, Meyer,
  Paavilainen, Olsson, and Persson}}]{ReppScience06}
\bibinfo{author}{\bibfnamefont{J.}~\bibnamefont{Repp}},
  \bibinfo{author}{\bibfnamefont{G.}~\bibnamefont{Meyer}},
  \bibinfo{author}{\bibfnamefont{S.}~\bibnamefont{Paavilainen}},
  \bibinfo{author}{\bibfnamefont{F.~E.} \bibnamefont{Olsson}},
  \bibnamefont{and} \bibinfo{author}{\bibfnamefont{M.}~\bibnamefont{Persson}},
  \bibinfo{journal}{Science} \textbf{\bibinfo{volume}{312}},
  \bibinfo{pages}{1196} (\bibinfo{year}{2006}).

\bibitem[{\citenamefont{Freund}(2002)}]{FreundReview}
\bibinfo{author}{\bibfnamefont{H.-J.} \bibnamefont{Freund}},
  \bibinfo{journal}{Surf. Sci.} \textbf{\bibinfo{volume}{500}},
  \bibinfo{pages}{271} (\bibinfo{year}{2002}).

\bibitem[{\citenamefont{Goodman}(2002)}]{GoodmanReview}
\bibinfo{author}{\bibfnamefont{D.}~\bibnamefont{Goodman}}, \bibinfo{journal}{J.
  Catal.} \textbf{\bibinfo{volume}{216}}, \bibinfo{pages}{213}
  (\bibinfo{year}{2002}).

\bibitem[{\citenamefont{Aulbur et~al.}(2000)\citenamefont{Aulbur, J\"onsson,
  and Wilkins}}]{Aulbur}
\bibinfo{author}{\bibfnamefont{W.~G.} \bibnamefont{Aulbur}},
  \bibinfo{author}{\bibfnamefont{L.}~\bibnamefont{J\"onsson}},
  \bibnamefont{and} \bibinfo{author}{\bibfnamefont{J.~W.}
  \bibnamefont{Wilkins}}, \bibinfo{journal}{Solid State Phys.: Advances in
  Research and Applications} \textbf{\bibinfo{volume}{54}}, \bibinfo{pages}{1}
  (\bibinfo{year}{2000}).

\bibitem[{\citenamefont{Rinke et~al.}(2005)\citenamefont{Rinke, Qteish,
  Neugebauer, Freysoldt, and Scheffler}}]{Rinke}
\bibinfo{author}{\bibfnamefont{P.}~\bibnamefont{Rinke}},
  \bibinfo{author}{\bibfnamefont{A.}~\bibnamefont{Qteish}},
  \bibinfo{author}{\bibfnamefont{J.}~\bibnamefont{Neugebauer}},
  \bibinfo{author}{\bibfnamefont{C.}~\bibnamefont{Freysoldt}},
  \bibnamefont{and}
  \bibinfo{author}{\bibfnamefont{M.}~\bibnamefont{Scheffler}},
  \bibinfo{journal}{New J. Phys.} \textbf{\bibinfo{volume}{7}},
  \bibinfo{pages}{126} (\bibinfo{year}{2005}).

\bibitem[{\citenamefont{F\"olsch et~al.}(1989)\citenamefont{F\"olsch,
  Barjenbruch, and Henzler}}]{Foelsch}
\bibinfo{author}{\bibfnamefont{S.}~\bibnamefont{F\"olsch}},
  \bibinfo{author}{\bibfnamefont{U.}~\bibnamefont{Barjenbruch}},
  \bibnamefont{and} \bibinfo{author}{\bibfnamefont{M.}~\bibnamefont{Henzler}},
  \bibinfo{journal}{Thin Solid Films} \textbf{\bibinfo{volume}{172}},
  \bibinfo{pages}{123} (\bibinfo{year}{1989}).

\bibitem[{\citenamefont{Zielasek et~al.}(2000)\citenamefont{Zielasek,
  Hildebrandt, and Henzler}}]{Zielasek00}
\bibinfo{author}{\bibfnamefont{V.}~\bibnamefont{Zielasek}},
  \bibinfo{author}{\bibfnamefont{T.}~\bibnamefont{Hildebrandt}},
  \bibnamefont{and} \bibinfo{author}{\bibfnamefont{M.}~\bibnamefont{Henzler}},
  \bibinfo{journal}{Phys. Rev. B} \textbf{\bibinfo{volume}{62}},
  \bibinfo{pages}{2912} (\bibinfo{year}{2000}).

\bibitem[{\citenamefont{Olsson et~al.}(2005)\citenamefont{Olsson, Persson,
  Repp, and Meyer}}]{Olsson}
\bibinfo{author}{\bibfnamefont{F.~E.} \bibnamefont{Olsson}},
  \bibinfo{author}{\bibfnamefont{M.}~\bibnamefont{Persson}},
  \bibinfo{author}{\bibfnamefont{J.}~\bibnamefont{Repp}}, \bibnamefont{and}
  \bibinfo{author}{\bibfnamefont{G.}~\bibnamefont{Meyer}},
  \bibinfo{journal}{Phys. Rev. B} \textbf{\bibinfo{volume}{71}},
  \bibinfo{pages}{075419} (\bibinfo{year}{2005}).

\bibitem[{\citenamefont{Schwennicke et~al.}(1993)\citenamefont{Schwennicke,
  Schimmelpfennig, and Pfn\"ur}}]{Schwennicke}
\bibinfo{author}{\bibfnamefont{C.}~\bibnamefont{Schwennicke}},
  \bibinfo{author}{\bibfnamefont{J.}~\bibnamefont{Schimmelpfennig}},
  \bibnamefont{and} \bibinfo{author}{\bibfnamefont{H.}~\bibnamefont{Pfn\"ur}},
  \bibinfo{journal}{Surf. Sci.} \textbf{\bibinfo{volume}{293}},
  \bibinfo{pages}{57} (\bibinfo{year}{1993}).

\bibitem[{\citenamefont{Gl\"ockler et~al.}(1996)\citenamefont{Gl\"ockler,
  Sokolowski, Soukopp, and Umbach}}]{Gloeckler}
\bibinfo{author}{\bibfnamefont{K.}~\bibnamefont{Gl\"ockler}},
  \bibinfo{author}{\bibfnamefont{M.}~\bibnamefont{Sokolowski}},
  \bibinfo{author}{\bibfnamefont{A.}~\bibnamefont{Soukopp}}, \bibnamefont{and}
  \bibinfo{author}{\bibfnamefont{E.}~\bibnamefont{Umbach}},
  \bibinfo{journal}{Phys. Rev. B} \textbf{\bibinfo{volume}{54}},
  \bibinfo{pages}{7705} (\bibinfo{year}{1996}).

\bibitem[{\citenamefont{Zielasek et~al.}(2004)\citenamefont{Zielasek,
  Hildebrandt, and Henzler}}]{Zielasek}
\bibinfo{author}{\bibfnamefont{V.}~\bibnamefont{Zielasek}},
  \bibinfo{author}{\bibfnamefont{T.}~\bibnamefont{Hildebrandt}},
  \bibnamefont{and} \bibinfo{author}{\bibfnamefont{M.}~\bibnamefont{Henzler}},
  \bibinfo{journal}{Phys. Rev. B} \textbf{\bibinfo{volume}{69}},
  \bibinfo{pages}{205313} (\bibinfo{year}{2004}).

\bibitem[{\citenamefont{White et~al.}(1997)\citenamefont{White, Godby, Rieger,
  and Needs}}]{White/Godby/Rieger/Needs:1997}
\bibinfo{author}{\bibfnamefont{I.~D.} \bibnamefont{White}},
  \bibinfo{author}{\bibfnamefont{R.~W.} \bibnamefont{Godby}},
  \bibinfo{author}{\bibfnamefont{M.~M.} \bibnamefont{Rieger}},
  \bibnamefont{and} \bibinfo{author}{\bibfnamefont{R.~J.} \bibnamefont{Needs}},
  \bibinfo{journal}{Phys.\ Rev.\ Lett.} \textbf{\bibinfo{volume}{80}},
  \bibinfo{pages}{4265} (\bibinfo{year}{1998}).

\bibitem[{\citenamefont{Fratesi et~al.}(2003)\citenamefont{Fratesi, Brivio,
  Rinke, and Godby}}]{Fratesi1}
\bibinfo{author}{\bibfnamefont{G.}~\bibnamefont{Fratesi}},
  \bibinfo{author}{\bibfnamefont{G.~P.} \bibnamefont{Brivio}},
  \bibinfo{author}{\bibfnamefont{P.}~\bibnamefont{Rinke}}, \bibnamefont{and}
  \bibinfo{author}{\bibfnamefont{R.~W.} \bibnamefont{Godby}},
  \bibinfo{journal}{Phys.\ Rev.\ B} \textbf{\bibinfo{volume}{68}},
  \bibinfo{pages}{195404} (\bibinfo{year}{2003}).

\bibitem[{\citenamefont{Rohlfing et~al.}(2003)\citenamefont{Rohlfing, Wang,
  Kr\"uger, and Pollmann}}]{Rohlfing/Wang/Krueger/Pollmann:2003}
\bibinfo{author}{\bibfnamefont{M.}~\bibnamefont{Rohlfing}},
  \bibinfo{author}{\bibfnamefont{N.-P.} \bibnamefont{Wang}},
  \bibinfo{author}{\bibfnamefont{P.}~\bibnamefont{Kr\"uger}}, \bibnamefont{and}
  \bibinfo{author}{\bibfnamefont{J.}~\bibnamefont{Pollmann}},
  \bibinfo{journal}{Phys.\ Rev.\ Lett.} \textbf{\bibinfo{volume}{91}},
  \bibinfo{pages}{256802} (\bibinfo{year}{2003}).

\bibitem[{\citenamefont{Rinke et~al.}(2004)\citenamefont{Rinke, Delaney,
  Garc\'{\i}a-Gonz\'alez, and Godby}}]{ClusterImStates:2004}
\bibinfo{author}{\bibfnamefont{P.}~\bibnamefont{Rinke}},
  \bibinfo{author}{\bibfnamefont{K.}~\bibnamefont{Delaney}},
  \bibinfo{author}{\bibfnamefont{P.}~\bibnamefont{Garc\'{\i}a-Gonz\'alez}},
  \bibnamefont{and} \bibinfo{author}{\bibfnamefont{R.~W.} \bibnamefont{Godby}},
  \bibinfo{journal}{Phys.\ Rev.\ A} \textbf{\bibinfo{volume}{70}},
  \bibinfo{pages}{063201} (\bibinfo{year}{2004}).

\bibitem[{SFH()}]{SFHIngX}
\bibinfo{note}{{}http://www.sphinxlib.de}.

\bibitem[{Siz()}]{SizeComment}
\bibinfo{note}{The 2$\times$1 surface cell has 2 atoms per Ge layer (4 per NaCl
  layer). A vacuum region of $\sim$10\,\AA{} proved sufficient to decouple the
  slabs. While bulk Ge is a metal in LDA, the 6-layer slab has a gap of 0.2\,eV
  due to quantum confinement, which simplifies the subsequent $G_0W_0$
  calculations.}

\bibitem[{\citenamefont{Rieger et~al.}(1999)}]{RiegerEtal}
\bibinfo{author}{\bibfnamefont{M.}~\bibnamefont{Rieger}} \bibnamefont{et~al.},
  \bibinfo{journal}{Comput.\ Phys.\ Commun.} \textbf{\bibinfo{volume}{117}},
  \bibinfo{pages}{211} (\bibinfo{year}{1999}).

\bibitem[{\citenamefont{Steinbeck et~al.}(2000)}]{gwstGLEtal}
\bibinfo{author}{\bibfnamefont{L.}~\bibnamefont{Steinbeck}}
  \bibnamefont{et~al.}, \bibinfo{journal}{Comput.\ Phys.\ Commun.}
  \textbf{\bibinfo{volume}{125}}, \bibinfo{pages}{105} (\bibinfo{year}{2000}).

\bibitem[{\citenamefont{Freysoldt et~al.}(2007{\natexlab{b}})}]{gwstAnisoEtal}
\bibinfo{author}{\bibfnamefont{C.}~\bibnamefont{Freysoldt}}
  \bibnamefont{et~al.}, \bibinfo{journal}{Comput. Phys. Commun.}
  \textbf{\bibinfo{volume}{176}}, \bibinfo{pages}{1}
  (\bibinfo{year}{2007}{\natexlab{b}}).

\bibitem[{\citenamefont{Freysoldt et~al.}(2008)\citenamefont{Freysoldt, Eggert,
  Rinke, Schindlmayr, and Scheff\-ler}}]{slabgw}
\bibinfo{author}{\bibfnamefont{C.}~\bibnamefont{Freysoldt}},
  \bibinfo{author}{\bibfnamefont{P.}~\bibnamefont{Eggert}},
  \bibinfo{author}{\bibfnamefont{P.}~\bibnamefont{Rinke}},
  \bibinfo{author}{\bibfnamefont{A.}~\bibnamefont{Schindlmayr}},
  \bibnamefont{and}
  \bibinfo{author}{\bibfnamefont{M.}~\bibnamefont{Scheff\-ler}},
  \bibinfo{journal}{Phys. Rev. B} \textbf{\bibinfo{volume}{77}},
  \bibinfo{pages}{235428} (\bibinfo{year}{2008}).

\bibitem[{\citenamefont{Rohlfing}(2000)}]{RohlfingMgO}
\bibinfo{author}{\bibfnamefont{M.}~\bibnamefont{Rohlfing}},
  \bibinfo{journal}{Int. J. Quantum Chem.} \textbf{\bibinfo{volume}{80}},
  \bibinfo{pages}{807} (\bibinfo{year}{2000}).

\bibitem[{\citenamefont{Neaton et~al.}(2006)\citenamefont{Neaton, Hybertsen,
  and Louie}}]{benzeneGraphite}
\bibinfo{author}{\bibfnamefont{J.~B.} \bibnamefont{Neaton}},
  \bibinfo{author}{\bibfnamefont{M.~S.} \bibnamefont{Hybertsen}},
  \bibnamefont{and} \bibinfo{author}{\bibfnamefont{S.~G.} \bibnamefont{Louie}},
  \bibinfo{journal}{Phys. Rev. Lett.} \textbf{\bibinfo{volume}{97}},
  \bibinfo{pages}{216405} (\bibinfo{year}{2006}).

\bibitem[{\citenamefont{Thygesen and Rubio}(2009)}]{Thygesen/Rubio:2009}
\bibinfo{author}{\bibfnamefont{K.~S.} \bibnamefont{Thygesen}} \bibnamefont{and}
  \bibinfo{author}{\bibfnamefont{A.}~\bibnamefont{Rubio}},
  \bibinfo{journal}{Phys.\ Rev.\ Lett.} \textbf{\bibinfo{volume}{102}},
  \bibinfo{pages}{046802} (\bibinfo{year}{2009}).

\bibitem[{foo()}]{footnote:VBstates}
\bibinfo{note}{Since a meaningful analysis of the NaCl's conduction bands is
  aggravated by their very strong hybridisation with Ge states, we will focus
  on the valence states only.}

\bibitem[{\citenamefont{Barjenbruch et~al.}(1989)\citenamefont{Barjenbruch,
  F\"olsch, and Henzler}}]{Barjenbruch}
\bibinfo{author}{\bibfnamefont{U.}~\bibnamefont{Barjenbruch}},
  \bibinfo{author}{\bibfnamefont{S.}~\bibnamefont{F\"olsch}}, \bibnamefont{and}
  \bibinfo{author}{\bibfnamefont{M.}~\bibnamefont{Henzler}},
  \bibinfo{journal}{Surf. Sci.} \textbf{\bibinfo{volume}{211/212}},
  \bibinfo{pages}{749} (\bibinfo{year}{1989}).

\end{thebibliography}

\end{document}